\documentstyle[12pt]{article}

\input{epsf}

\def\bsgam{b\rightarrow s\gamma}

\def\be{\begin{equation}}
\def\ee{\end{equation}}
\def\bea{\begin{eqnarray}}
\def\eea{\end{eqnarray}}

\def\g2{$g_\mu-2$}

\begin{document}

\begin{titlepage}

\makebox{ }
\vspace{ 0.1in}

\begin{flushright}
        NUHEP-99-81\\
        December 1999
\end{flushright}

\vskip 1in

\begin{center}

{\large \bf 
Muon $g-2$ in the MSSM constrained by simple SO(10) SUSY GUT
}

\vskip 0.4in

Tom\'{a}\v{s} Bla\v{z}ek$^{*}$

\vskip 0.2in

{\em Department of Physics and Astronomy,
     Northwestern University,
     Evanston, IL 60208 }\\

\vskip 0.2in

E-mail: {\em blazek@heppc19.phys.nwu.edu}

\end{center}

\vskip 0.7in

\begin{abstract}

We show that the best fits of the MSSM constrained by a simple SO(10) SUSY GUT
are consistent with the present data on the muon anomalous magnetic moment. 
The best fits assume rather large values of tan$\beta\approx 50$, and in our analysis 
they are not {\em a priori} correlated in any way, directly or indirectly, 
with the experimental limit on $a_\mu$.
Regions in the SUSY parameter space, 
which are currently ruled out because of too large $a_\mu^{SUSY}$, 
are already excluded in the global fit by excessive corrections to $m_b$, unacceptable 
$BR(b\rightarrow s\gamma)$, or direct experimental limits on sparticle
masses. However, our results indicate that the accuracy expected in the ongoing E821 
experiment at BNL will eventually turn the muon anomalous magnetic moment into a major
constraint for this regime of the MSSM.

\end{abstract}

\vskip 1.0in


\vskip 0.1in

$^*${\footnotesize On leave of absence from 
the Dept. of Theoretical Physics, Comenius Univ., Bratislava, Slovakia}

\end{titlepage}

\renewcommand{\thepage}{\arabic{page}}
\setcounter{page}{1}

\section{Introduction}

The anomalous magnetic moment of the muon, $a_\mu\equiv (g_\mu-2)/2$, is potentially a significant 
constraint for any extension of the Standard Model (SM). The measured value quoted by the 
Particle Data Group \cite{a_mu_PDG} is
\be
a_\mu^{PDG} = (11\,659\,230 \pm 84) \times 10^{-10}.
\ee
This value does not take into account the early results \footnote{The 1998 result is still preliminary.}
from the E821 experiment at BNL 
\cite{a_mu_BNL97,a_mu_BNL98}
\bea
 & a_\mu^{BNL\,97} = (11\,659\,250 \pm 150) \times 10^{-10}, \\
 & a_\mu^{BNL\,98} = (11\,659\,191 \pm \,59) \times 10^{-10}. 
\eea
On the other hand, the SM prediction yields (\cite{a_mu_SM} and references therein) 
\be
a_\mu^{SM} = (11\,659\,159.6 \pm 6.7) \times 10^{-10},
\label{a_SM}
\ee
where the QED, electroweak and hadronic contributions are sumed and the errors are combined
in quadrature. When we take into account all these results, contributions from new physics 
beyond the SM are constrained to fit within the window
\be
    -28 \times 10^{-10} < \delta a_\mu^{NEW} < +124 \times 10^{-10}, \;\;  {\rm at}\; 90\%\: C.L.
\label{da_now}
\ee
The width of the window is at present dominated by the experimental
uncertainty. That, however, will change after the E821 experiment is completed.
The inclusion of the current BNL data already reduces the available window by a factor of 2.
After two more years of running the E821 is eventually expected to reach the accuracy $\pm 3 \times 10^{-10}$. 
If the measured central value then turns out to be exactly equal to the SM prediction, eq.(\ref{a_SM}), 
the new constraint will be 
\be
    -12 \times 10^{-10} < \delta a_\mu^{NEW} < +12 \times 10^{-10}, \;\;  {\rm at}\; 90\%\: C.L.,
\label{da_99}
\ee
with the window for new physics narrowed by more than a factor of 6. \footnote
{
This estimate does not take into 
account possible improvements over time in the
hadronic uncertainty, and thus the 
constraint on new physics may actually
be even tighter.
}

In the MSSM, there are significant contributions from new physics due to the chargino-sneutrino 
and neutralino-smuon loops: 
\bea
\delta a_\mu^{\chi^+} &=& \frac{1}{8\pi^2}\;
                           \sum_{A \alpha}\: \frac{m_\mu^2}{m_{\tilde{\nu}_\alpha}^2}
                           \left[ (|C_{A\alpha}^L|^2 + |C_{A\alpha}^R|^2)\: F_1(x_{A\alpha})
            + \frac{m_{\chi^+_A}}{m_\mu}\, {Re}\{ C_{A\alpha}^L\, C_{A\alpha}^{R*} \}\:F_3(x_{A\alpha})
                           \right]\: ,
\label{eq:daC}
\\
\delta a_\mu^{\chi^0} &=& - \frac{1}{8\pi^2}\;
                           \sum_{a \alpha}\: \frac{m_\mu^2}{m_{\tilde{\mu}_\alpha}^2}
                           \left[ (|N_{a\alpha}^L|^2 + |N_{a\alpha}^R|^2)\: F_2(x_{a\alpha})
            + \frac{m_{\chi^0_a}}{m_\mu}\, {Re}\{ N_{a\alpha}^L\, N_{a\alpha}^{R*} \}\: F_4(x_{a\alpha})
                           \right]\: ,
\label{eq:daN}
\eea
where $x_{A\alpha} = m_{\chi^+_A}^2 / m_{\tilde{\nu}_\alpha}^2$, 
   $\; x_{a\alpha} = m_{\chi^0_a}^2 / m_{\tilde{\mu}_\alpha}^2$, 
\bea
      C_{A\alpha}^L &=& - g_2       \, V_{A1}\, (\Gamma_{\nu L}^\dagger)_{2\alpha}\: ,\nonumber \\
      C_{A\alpha}^R &=& +\lambda_\mu\, U_{A2}\, (\Gamma_{\nu L}^\dagger)_{2\alpha}\: ,\nonumber \\
      N_{a\alpha}^L &=& -\lambda_\mu\, N_{a3}\, (\Gamma_{e   R}^\dagger)_{2\alpha}
                      +(\frac{g_1}{\sqrt{2}} N_{a1} + \frac{g_2}{\sqrt{2}} N_{a2} )
                                             \, (\Gamma_{e   L}^\dagger)_{2\alpha}\: ,\nonumber \\
      N_{a\alpha}^R &=& -\lambda_\mu\, N_{a3}\, (\Gamma_{e   L}^\dagger)_{2\alpha}
                        -\sqrt{2}\:g_1 N_{a1}\, (\Gamma_{e   R}^\dagger)_{2\alpha}\: ,
\eea
and the standard integrals over the two-dimensional Feynman parameter space are
\bea
     F_1(x) &=& \frac{1}{12(x-1)^4}\: ( x^3-6x^2+3x+2+6x  \,\ln x)\: ,\nonumber \\
     F_2(x) &=& \frac{1}{12(x-1)^4}\: (2x^3+3x^2-6x+1-6x^2\,\ln x)\: ,\nonumber \\
     F_3(x) &=& \frac{1}{ 2(x-1)^3}\: (      x^2-4x+3+2   \,\ln x)\: ,\nonumber \\
     F_4(x) &=& \frac{1}{ 2(x-1)^3}\: (      x^2   -1-2x  \,\ln x)\: .
\eea
In these relations, notation of \cite{bcrw,hk,bPhD} is assumed.
$C$'s and $N$'s are the couplings of the chiral muon states with the charginos and neutralinos, respectively.
$U_{A2}$, $V_{A1}$ and $N_{a3}$ are the elements of the chargino and neutralino mixing matrices, 
$A=1,2$; $a=1,4$. $\Gamma$'s are the mixing matrices of the sleptons after the flavor
eigenstates have been first rotated by the unitary matrices which diagonalize the respective fermionic states. 
Note that $\alpha=1,3$ for sneutrinos and $\alpha=1,6$ for charged sleptons. In particular, for the charged sleptons
each $(\Gamma_e)_{\alpha i}$ is a 6$\times$3 matrix defined as 
$(\Gamma_{eL})_{\alpha i}=Z_{\alpha i}$,  $(\Gamma_{eR})_{\alpha i}=Z_{\alpha\: i+3}$, ($i=1,3$), where
$Z$ is a 6$\times$6 mixing matrix for charged sleptons of all three generations.
For 3 sneutrinos, $\Gamma_{\nu L}$ directly diagonalizes the 3$\times$3 
sneutrino mass matrix.
$m_{\chi^+_A}$, $m_{\chi^0_a}$, $m_{\tilde{\nu}_\alpha}^2$ and $m_{\tilde{\mu}_\alpha}^2$
are the chargino, neutralino, sneutrino and charged slepton mass eigenvalues, $m_\mu$ and $\lambda_\mu$ are 
the muon mass and diagonal muon Yukawa coupling, and 
$g_1$ and $g_2$ are the electroweak gauge couplings.

Note that at first glance the terms proportional to $F_3$ and $F_4$ in eqs.(\ref{eq:daC}, \ref{eq:daN}), 
respectively, are enhanced by $m_\chi/m_\mu$. The net enahncement is actually of the order $v_u/v_d \equiv \tan\!\beta$.
($v_d$ and $v_u$ are the Higgs vevs which give masses
to the $d$-quarks and charged leptons, and to the $u$-quarks, respectively.) \footnote
{
Note that there is no additional enhancement due to $1/\lambda_\mu$ in the terms proportional to $F_3$ and $F_4$ 
after the sum over the mass eigenstates is performed. In fact, any combination
of chiral muon states contributes to  $ \delta a_\mu^{SUSY} $
with the net contribution {\em suppressed} by small Yukawa coupling $\lambda_\mu$, as one can see, 
for instance, from the relevant Feynman diagrams in terms of flavor eigenstates ({\em i.e.}, 
in the interaction basis).
That also explains why the electron anomalous magnetic moment is less sensitive to new physics
than its muon analogy.
}
The enhancement by tan$\beta$ can be traced back to the diagrams with the chirality flip inside the loop
(or in one of the vertices) as opposed to the terms proportional to $F_1$ and $F_2$ where the chirality
flip takes place in the external muon leg. There are no similarly enhanced terms in the SM, 
where the chirality can only be flipped in the external muon leg. Thus for tan$\beta \gg 1$
we expect that the terms proportional to $F_3$ and $F_4$ dominate in eqs.(\ref{eq:daC}) and (\ref{eq:daN}),  
and that compared to the SM electroweak gauge boson contribution $ \delta a_\mu^{EW} $, the SUSY contribution
scales approximately as 
\be
 \delta a_\mu^{SUSY} \simeq   \delta a_\mu^{EW} \: \left( \frac{M_W          }{\tilde{m}} \right)^2\, \tan\!\beta
                     \simeq  15 \times 10^{-10} \: \left( \frac{100 {\rm GeV}}{\tilde{m}} \right)^2\, \tan\!\beta,
\label{eq:compEW}
\ee
where $\tilde{m}$ stands for the heaviest sparticle mass in the loop. The effect has been noticed and emphasized
by the earlier studies focusing on the muon anomalous magnetic moment in the context of the MSSM \cite{a_mu_MSSM}.

Relation (\ref{eq:compEW}) predicts a very simple tan$\beta$ dependence. It suggests that the large tan$\beta$ regime
of the MSSM may already be constrained by the currently allowed window, eq.(\ref{da_now}).
That is of interest for models of grand unification based on simple SO(10), since the SO(10)
GUT constraint $y_t=y_b=y_\tau$ for the Yukawa couplings of the third generation implies tan$\beta\approx 50$.
In this study, we first apply our analysis to the MSSM constrained only by gauge coupling unification 
and, as a warm-up, compute the muon anomalous magnetic moment for fixed values tan$\beta=2$ and tan$\beta=20$. 
Next, we proceed to the best fits of a simple SO(10) SUSY GUT \cite{lr}. We present our results 
in terms of the contour lines of constant $a_\mu^{SUSY}$ drawn in the SUSY parameter space. 
In section 2, we review our numerical analysis. Sections 3 and 4 discuss the results and 
prospects for the future.

\section{Numerical Analysis}

%

Our numerical analysis has relied on the top down global analysis introduced in \cite{bcrw}.
The soft SUSY breaking mediated by supergravity (SUGRA)\cite{an} was assumed throughout this paper. 
As explained at the end of the previous section, the magnitude of tan$\beta$ is of special interest for the analysis.
Hence we discuss separately three cases with low, medium, 
and large value of tan$\beta$. First, fixed tan$\beta=2$ was considered, along with
the scalar trilinear parameter $A_0$ set to zero to simplify the analysis which is insensitive to $A_0$
in this case. Next, tan$\beta$ was raised to $20$ and $A_0$ set free to vary. 
In the third case, we used the results of the global analysis of model 4c,
a simple SO(10) model with minimal number of effective operators leading to realistic Yukawa matrices. 
The model was suggested by Lucas and Raby \cite{lr} and its low-energy analysis was presented in \cite{bcrw}.
A direct model dependence of $a_\mu$ on the details of the Yukawa matrices is, however, very limited.
In this case, tan$\beta$ was not fixed to any particular value. 
Instead, it was a free parameter of the global analysis but 
since the model predicted exact $t-b-\tau$ Yukawa coupling unification, the best fit values of tan$\beta$
were always found between 50 and 55, dependent on a particular $(m_0,M_{1/2})$ point as explained
below. $A_0$ was free to vary --- as for tan$\beta=20$ --- which allowed the best fits to optimize the effects of 
the left-right stop mixing, which is enhanced by tan$\beta$, in the analysis.

In each of the three cases, gauge coupling unification was imposed up to a small (less than 5\%) 
negative correction, called $\epsilon_3$, to $\alpha_s$ at the unification scale $M_G$. 
Scale $M_G$ has been defined as the scale where $\alpha_1$ and $\alpha_2$ are exactly equal to
the common value $\alpha_G$.
For low and medium tan$\beta$ ($2$ and $20$), the minimal set of the initial SUSY parameters 
\be
M_{1/2},\; m_0,\; A_0,\; \mbox{\rm sign}\mu,\; \mbox{\rm and tan} \beta
\label{4params}
\ee
was assumed,
with $M_{1/2}$, $m_0$, and $A_0$ (the universal gaugino mass, scalar mass and trilinear coupling) 
introduced at $M_G$. The Yukawa couplings of the third generation
fermions at $M_G$ were unconstrained and free to vary independently on each other.
For large tan$\beta$ in the third case, 
the scalar Higgs masses were allowed to deviate from $m_0$
in order to alleviate strong fine-tuning required for the correct electroweak symmetry breaking.
Thus instead of (\ref{4params}), the set
\be
M_{1/2},\; m_0,\; (m_{H_d}/m_0)^2,\; (m_{H_u}/m_0)^2,\; 
A_0,\; \mbox{\rm sign}\mu,\; \mbox{\rm and tan} \beta
\label{6params}
\ee
was actually used as initial SUSY parameters. As already mentioned, the third generation yukawas were 
strictly set equal to each other at $M_G$ in this case. In fact, the SO(10)-based equality among them is 
the main reason why such a large tan$\beta$ is attractive.

The rest of the analysis was then the same for each of the three cases. 
Particular values of $m_0$ and $M_{1/2}$ were picked up, while the rest of the initial parameters varied.
That included varying $\alpha_G$, $M_G$, $\epsilon_3$,
the third generation Yukawa couplings, and $A_0$ (for medium and large tan$\beta$). 
Using the 2-loop RGEs for the dimensionless couplings
and 1-loop RGEs for the dimensionful couplings the theory was renormalized down to the SUSY scale, which was set equal
to the mass of the $Z$ boson. The electroweak symmetry breaking was checked to one loop as in \cite{bcrw},
based on the effective potential method of ref.\cite{EWSB}. One-loop SUSY threshold corrections
to fermion masses were calculated consistently at this scale. That is of particular importance for
$m_b$ which receives significant corrections if tan$\beta$ is large. Also, the experimental
constraints imposed by the observed branching ratio $BR(\bsgam)$ and by direct sparticle searches 
were taken into account.
Finally, $ \delta a_\mu^{SUSY} \equiv \delta a_\mu^{\chi^+} + \delta a_\mu^{\chi^0} $ 
was evaluated following eqs.(\ref{eq:daC}) and (\ref{eq:daN})
for those values of the initial parameters which gave the lowest $\chi^2$ calculated 
out of the ten low energy observables $M_Z$, $M_W$, $\rho_{new}$, $\alpha_s(M_Z)$, $\alpha$,
$G_\mu$, $M_t$, $m_b(M_b)$, $M_\tau$, and $BR(\bsgam)$. Details of the low energy analysis 
can be found in \cite{bcrw}. 
The calculated value of $ \delta a_\mu^{SUSY}$ did not have any effect on the 
$\chi^2$ calculation and the subsequent selection of the SUSY parameter space.

\section{Results and Discussion}

\begin{figure}[p]
\epsfysize=6truein
\epsffile{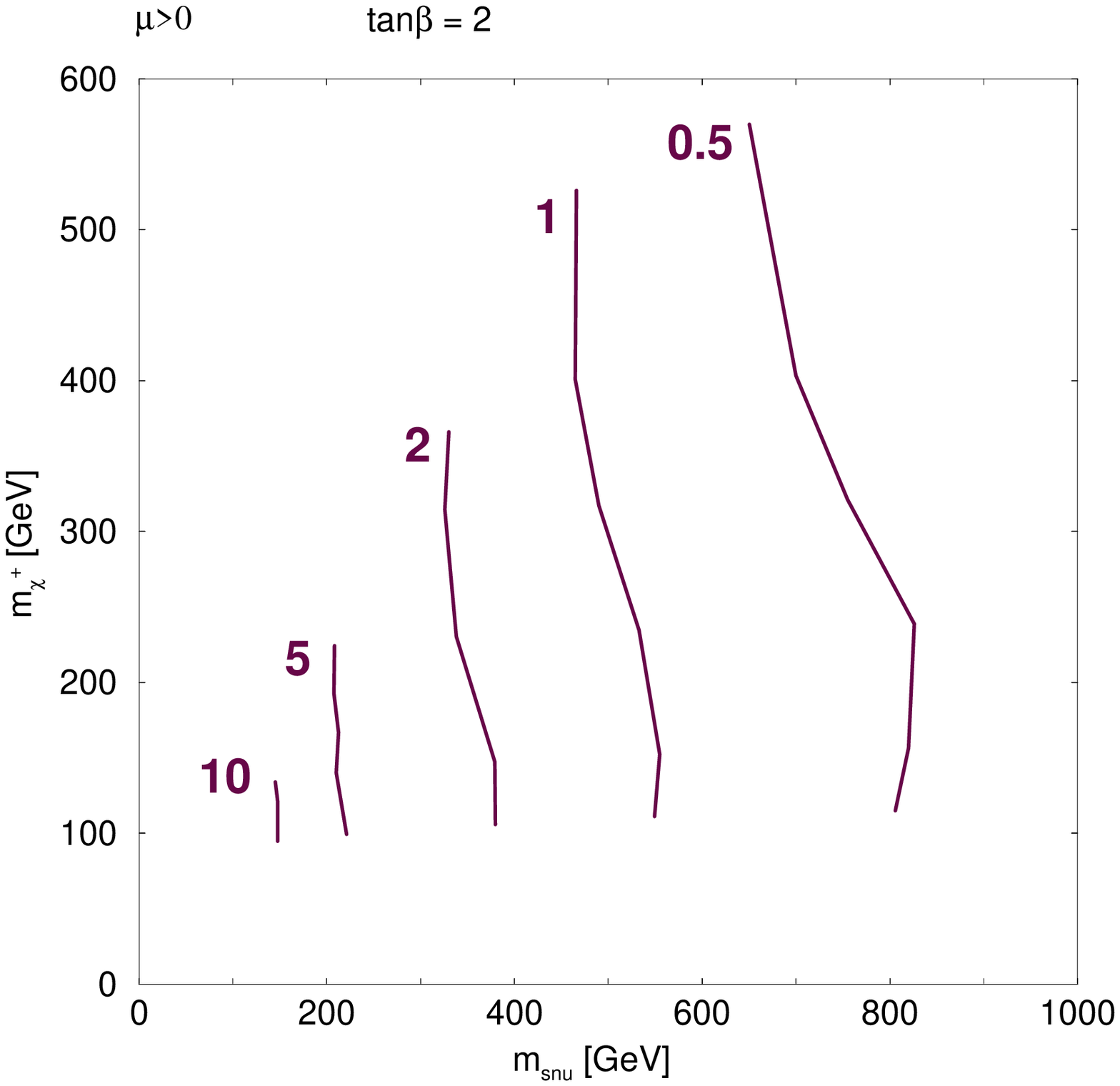}
\caption{
Contour lines of constant $\delta a_\mu^{SUSY}\times 10^{10}$ in the analysis
with fixed tan$\beta=2$, in the $(m_{\tilde{\nu}},m_{\chi^+})$ plane, where 
$m_{\tilde{\nu}}$ is the muon sneutrino mass and $m_{\chi^+}$ is the mass
of the lighter chargino.
}
\label{famu_tb2_mm}
\end{figure}

\begin{figure}[p]
\epsfysize=6truein
\epsffile{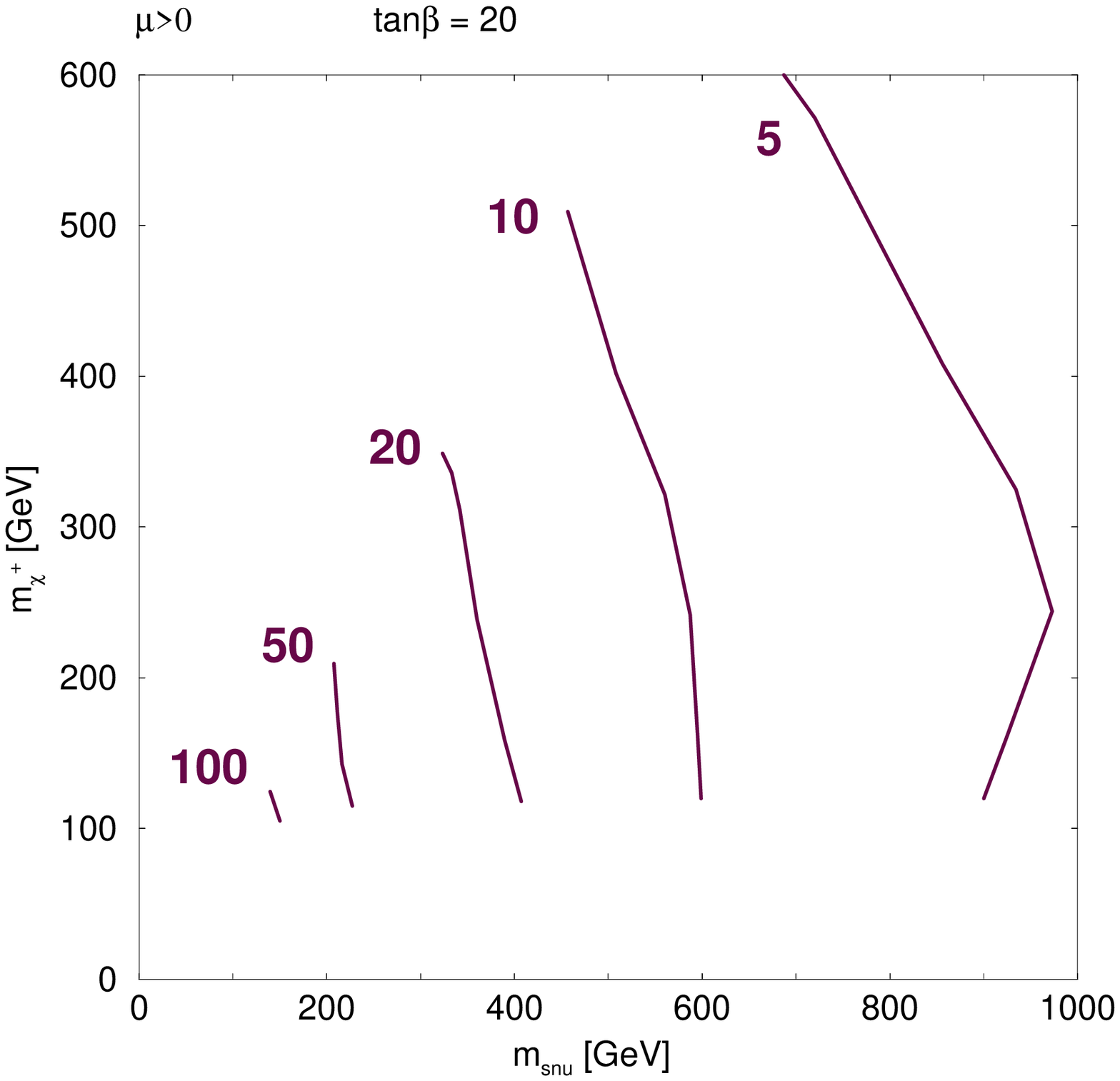}
\caption{
Contour lines of constant $\delta a_\mu^{SUSY}\times 10^{10}$ in the analysis
with fixed tan$\beta=20$, in the $(m_{\tilde{\nu}},m_{\chi^+})$ plane, where 
$m_{\tilde{\nu}}$ is the muon sneutrino mass and $m_{\chi^+}$ is the mass
of the lighter chargino.
}
\label{famu_tb20_mm}
\end{figure}

The results for tan$\beta=2$ and tan$\beta=20$ are shown in figures \ref{famu_tb2_SUSY}--\ref{famu_tb20_mm}. 
The figures \ref{famu_tb2_SUSY} and \ref{famu_tb20_SUSY} show the
contour lines of constant $ \delta a_\mu^{SUSY}\times 10^{10} $ in the $(m_0,M_{1/2})$ plane.
These results are then transformed into the dependence on physical masses in figures 
\ref{famu_tb2_mm} and \ref{famu_tb20_mm}, where we chose the 
$(m_{\tilde{\nu}},m_{\chi^+})$ plane, with $m_{\tilde{\nu}}$ being the muon sneutrino mass 
and $m_{\chi^+}$ being the mass of the lighter chargino. 
The contour lines in these figures are bound from below
by the limit on the neutralino mass (a limit $m_{\chi^0} > 55$GeV was imposed flatly, for simplicity),  
and from above by the stau mass  ($m_{\tilde{\tau}} > 60$GeV, in this analysis).
The important observation is that
the present limits on $ \delta a_\mu^{SUSY}$, eq.(\ref{da_now}) are not excluding
any region of the parameter space which is left available by other experiments. 
The size of $ \delta a_\mu^{SUSY} $ is in agreement with the results of Chattopadhyay and Nath, 
and of Moroi \cite{a_mu_MSSM}. We can also confirm their observation that the neutralino
contribution $\delta a_\mu^{\chi^0} $ is generally much smaller than the chargino contribution
$\delta a_\mu^{\chi^+} $ in the whole parameter space available. Yet several interesting characteristics
of the presented results cannot be read off directly from the earlier studies.
It is first of all the simple pattern suggested by
%
%
figures \ref{famu_tb2_mm} and \ref{famu_tb20_mm}.
The pattern in the figures shows a strong dependence of 
$ \delta a_\mu^{SUSY}$ on the sneutrino mass and very little sensitivity to the mass of the
lighter chargino. That can be understood from the fact that the loop integrals are in general
most sensitive to the heaviest mass in the loop, and the universal initial conditions (see (\ref{4params}))
together with the experimental limits from direct searches always lead to $m_{\tilde{\nu}}\geq m_{\chi^+}$
at the electroweak scale. 
%
%
Actually, the figures show how well the approximate relation (\ref{eq:compEW}) works. One could {\em e.g.},  
directly read out the value of sneutrino mass from a figure of this type once a more accurate measurement of $a_\mu$
and the value of tan$\beta$ become available.
Alternatively, a more accurate measurement of $a_\mu$ can be converted into a limit on 
tan$\beta$ in a straightforward way, if the sneutrino mass is known. Also note how well the linear dependence
$ \delta a_\mu^{SUSY} \propto \tan\!\beta$ holds. If we overlapped figures 
\ref{famu_tb2_mm} and \ref{famu_tb20_mm}, the
contour lines marked as 10, 5, 2, 1, and 0.5   in figure \ref{famu_tb2_mm}  would be practically on top of the
contour lines marked as 100, 50, 20, 10, and 5 in figure \ref{famu_tb20_mm}. That suggests that the 
tan$\beta$ enhanced terms in eqs.(\ref{eq:daC}) and (\ref{eq:daN}) become dominant already
for tan$\beta \geq 2$. We checked this feature in the numerical analysis: typically, the term proportional to $F_3$
in eq.(\ref{eq:daC}) accounts for about 80--85\% of the net $\delta a_\mu^{\chi^+}$ for tan$\beta=2$,
while it is 97-99\% for tan$\beta=20$. (The analogous term in eq.(\ref{eq:daN}) is less dominant, but
has a smaller net effect since $\delta a_\mu^{\chi^0} < \delta a_\mu^{\chi^+}$.) 

From these results one could extrapolate that the current $a_\mu$ limit, eq.(\ref{da_now}), places
an important constraint on the analysis if tan$\beta$ is as large as 50. 

\begin{figure}[p]
\epsfysize=6.0truein
\epsffile{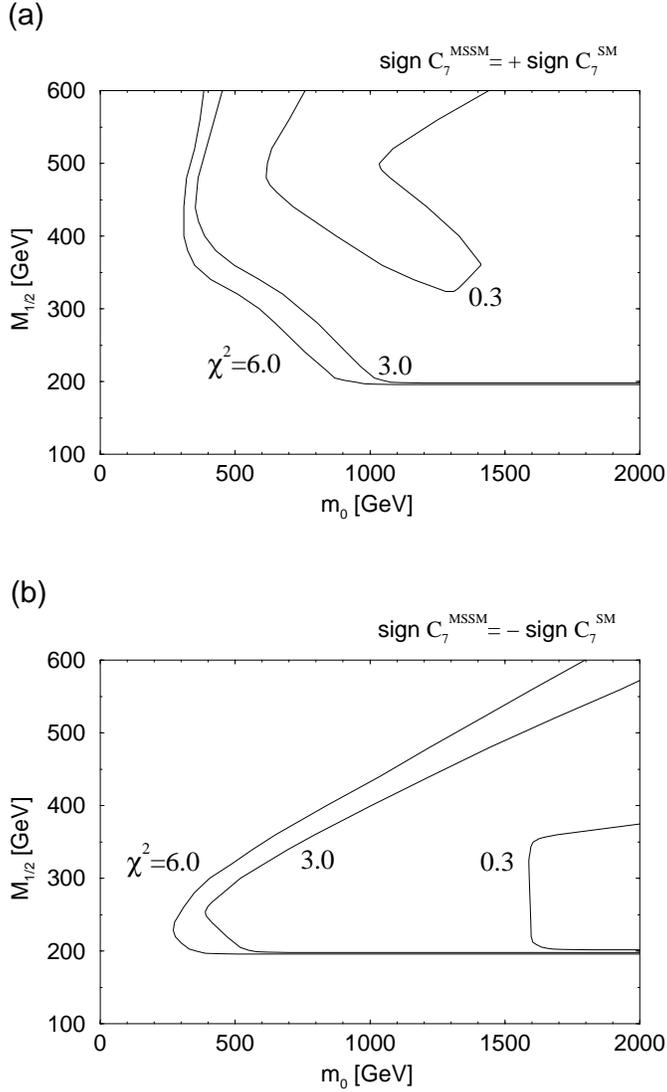}
\caption{
$\chi^2$ contour plots in the best fits of a simple SO(10) model, 
with the Wilson coefficient $C_7^{MSSM}$ (relevant for the $\bsgam$ decay) of 
({\em a}) the same ({\em b}) the opposite sign as compared to $C_7^{SM}$. 
As indicated, the contour lines correspond
to $\chi^2=6,\,3$, and $0.3$ per $3\; d.o.f.$, respectively.
tan$\beta$ varies between 50 and 55 due to the model prediction that 
$t$, $b$, and $\tau$ Yukawa couplings are equal at the GUT scale.
}
\label{f_F4c110}
\end{figure}

However, case tan$\beta\approx 50$ is qualitatively different.
As discussed in the study by Bla\v{z}ek and Raby \cite{bcrw}, 
the global analysis yields two distinct fits, figs. \ref{f_F4c110}a--b. 
The fits differ by the sign of the Wilson coefficient $C_7^{MSSM}$ in the effective quark Hamiltonian 
below the electroweak scale. (The coefficient $C_7$ determines the $\bsgam$ decay amplitude, after the QCD 
renormalization effects are taken into account \cite{misiak}.) 
In the MSSM with large tan$\beta$, the sign of $C_7$ can be the same, or the opposite, as in the SM. 
It can be reversed due to the fact that the chargino contribution can be of either sign. 
This contribution is enhanced by tan$\beta$ compared to the SM and charged Higgs contributions, whose
signs are fixed and alike, and thus the flipped sign of $C_7^{MSSM}$ cannot be obtained for low tan$\beta$. 
\footnote
{ 
The contributions with neutralinos or gluinos in the loop always turn out to be small in our analysis.
} 
For tan$\beta\approx 50$  the fit with the flipped sign is equally good as the fit with the sign 
unchanged, see figure \ref{f_F4c110}. The fits differ in the range of the allowed SUSY parameter space:  
to reverse the sign the chargino contribution accepts lower squark masses and different values
of $A_0$ than in the case when the sign is unchanged.

One can anticipate that these differences will be reflected in the analysis of $\delta a_\mu^{SUSY}$
with tan$\beta\approx 50$, 
since the quark sector is correlated with the lepton sector through the unification
constraint, and the soft masses are interrelated through the universality assumption.
Due to differences in SUSY spectra we can expect $\delta a_\mu^{SUSY}$ to be more significant in
the fit with the flipped sign of $C_7^{MSSM}$ than in the fit where the signs are the same.
These expectations are indeed realized in our results in figs. \ref{famu_F4c}a and \ref{famu_F4c}b.
In these figures, we plot the contour lines in the ($m_0,M_{1/2}$) plane of $\delta a_\mu^{SUSY}$, 
calculated according to eqs. (\ref{eq:daC}) and (\ref{eq:daN}), with all the masses, mixings 
and couplings taken over from the best fit values at the specific ($m_0,M_{1/2}$) point 
in the SUSY space.\footnote
{
We do not show plots in the $(m_{\tilde{\nu}},m_{\chi^+})$ plane in this case.
For large tan$\beta$ the best fits of the global analysis result in the lightest chargino
being higgsino-like, with its mass close to the electroweak scale across the whole 
$(m_{0},M_{1/2})$ plane. Thus the two-dimensional plots in the $(m_{\tilde{\nu}},m_{\chi^+})$ plane would
contract to a dependence on $m_{\tilde{\nu}}$ only.
}
The important feature, which runs contrary to the naive extrapolation from the previous two cases,
 is that the MSSM contribution to the muon anomalous magnetic moment {\em stays within the currently 
allowed range at $90\%\,C.L.$, even for tan$\beta$ as large as predicted by simple SO(10) GUTs}.
Despite the fact that figures \ref{famu_tb2_SUSY}-\ref{famu_tb20_mm} suggest that the analysis 
with tan$\beta = 50 - 55$ becomes sensitive to the current constraint on $a_\mu$, eq.(\ref{da_now}), 
our results in fig.\ref{famu_F4c} clearly indicate that this is not the case.
The reson for this is that we consistently demand that all known particle physics constraints 
(besides those imposed by $a_\mu$) are accounted for. 
For tan$\beta$ as large as 50, the allowed SUSY parameter range is further reduced by
strong constraints on the $b$ quark mass and the branching ratio $BR(\bsgam)$, when compared to
the regime where tan$\beta=2$ or $20$.

Finally, we make a note on the sign of $\mu$. As clearly indicated on the top of each figure, 
we have presented our results
just for $\mu > 0$. For this sign of the $\mu$ parameter, the SUSY contributions to $a_\mu$ are positive 
across the whole ($m_0,M_{1/2}$) plane. For $\mu < 0$, the contributions change sign too. However,
the chargino contribution to $C_7$ also changes sign in this case, which leads to unacceptable
values of $BR(\bsgam)$ for medium and large tan$\beta$. 
It is amazing to observe that for this range of tan$\beta$ the sign of the $\mu$ parameter 
favored by $\bsgam$ is the same as the sign preferred by the experimental window open for $\delta a_\mu^{NEW}$. 
\footnote
{
For low tan$\beta$, the $\bsgam$ constraint disappears
(the chargino contribution to $C_7$ is small) and $\mu$ can be negative. 
However, figure \ref{famu_tb2_SUSY} shows that $\delta a_\mu^{SUSY}$ 
is very small in this case. We know from the earlier studies \cite{a_mu_MSSM} that no substantial change
in the magnitude of $\delta a_\mu^{SUSY}$ occurs for different signs of $\mu$. Thus we conclude that 
for $\mu < 0$ only the low tan$\beta$ case is viable, but then the SUSY contributions to $a_\mu$ stay
below the electroweak contributions and will be hard to observe even after the E821 experiment is 
completed.
}

\section{Prospects for the New BNL Experiment and Conclusions}

\begin{figure}[p]
\epsfysize=6truein
\epsffile{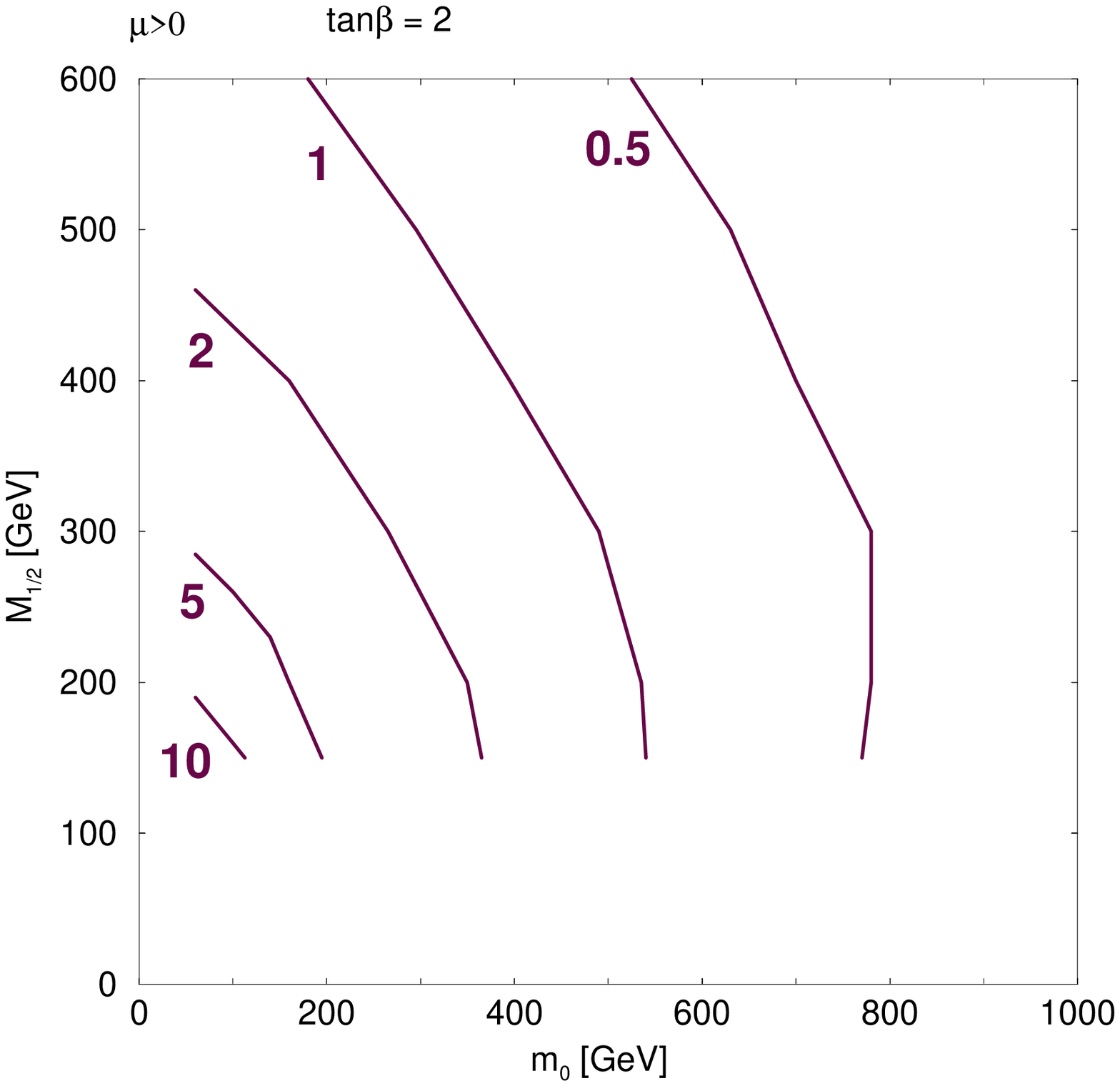}
\caption{
Contour lines of constant $\delta a_\mu^{SUSY}\times 10^{10}$ in the analysis
with fixed tan$\beta=2$ in the SUSY parameter space. 
}
\label{famu_tb2_SUSY}
\end{figure}

\begin{figure}[p]
\epsfysize=6truein
\epsffile{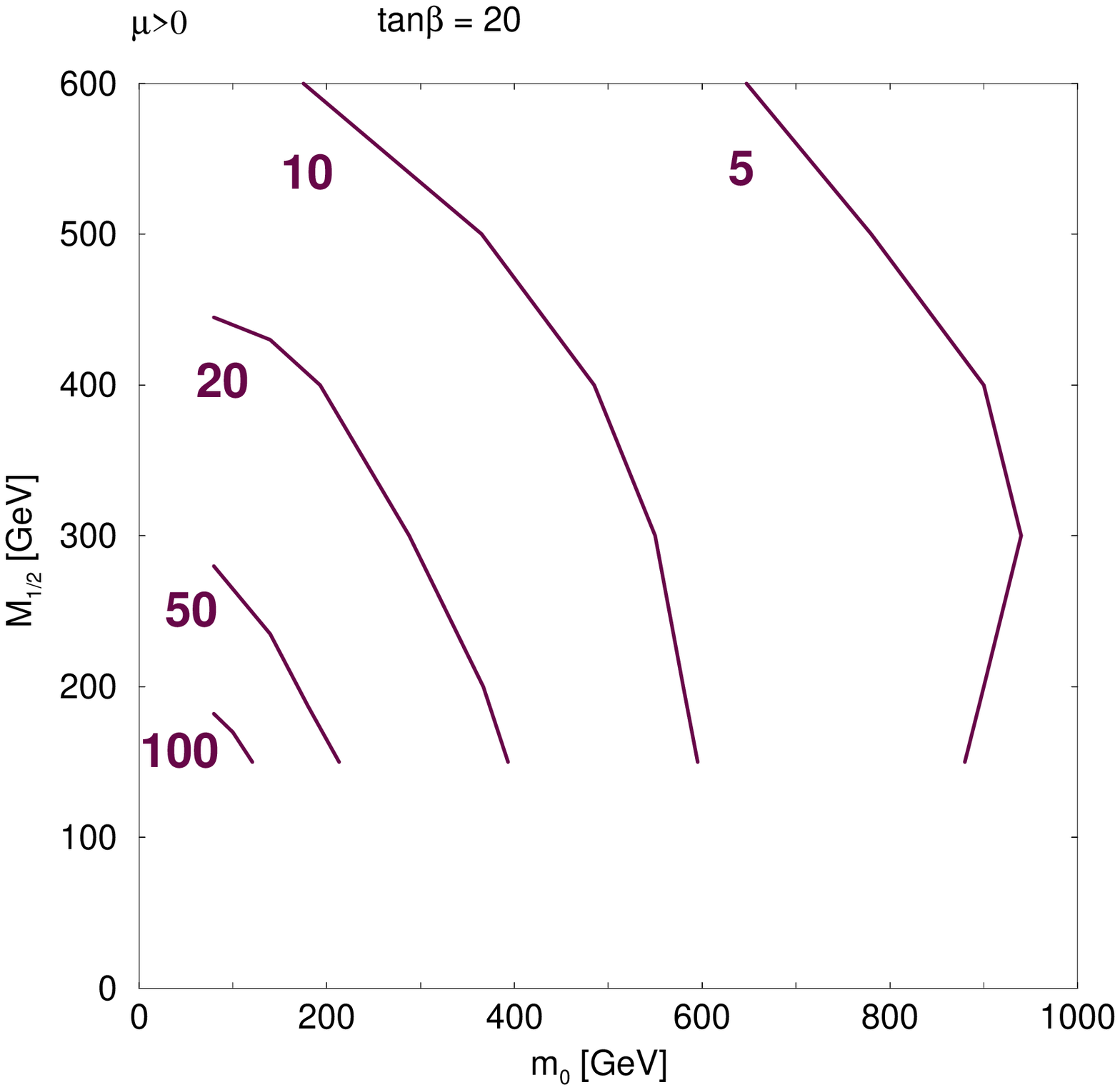}
\caption{
Contour lines of constant $\delta a_\mu^{SUSY}\times 10^{10}$ in the analysis
with fixed tan$\beta=20$, in the SUSY parameter space.
}
\label{famu_tb20_SUSY}
\end{figure}

\begin{figure}[p]
\epsfysize=6truein
\epsffile{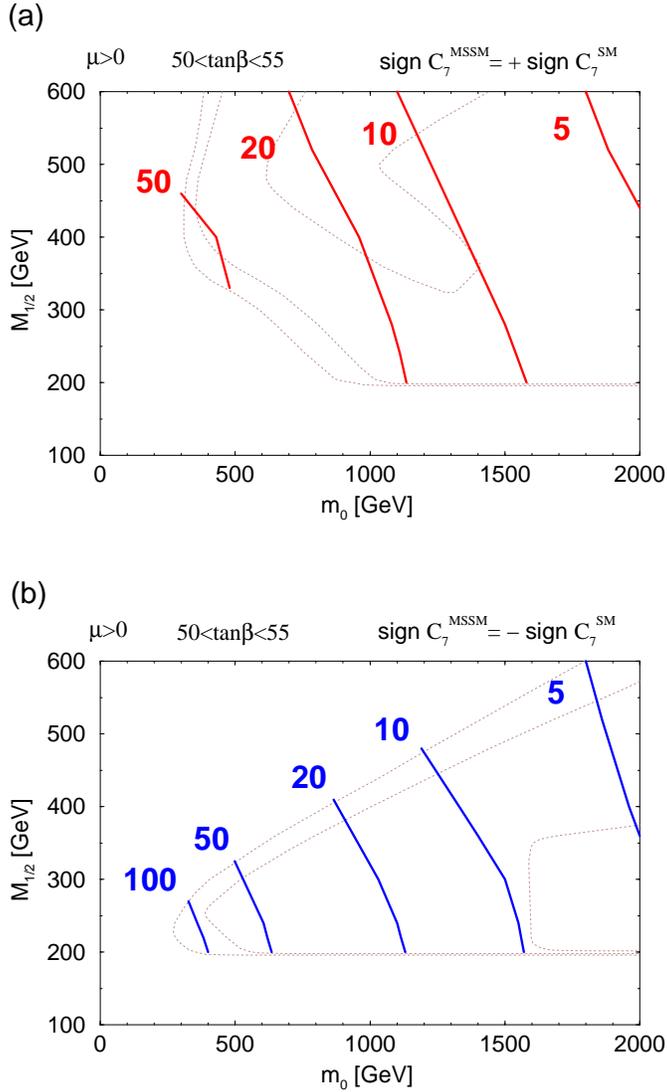}
\caption{
Contour lines of constant $\delta a_\mu^{SUSY}\times 10^{10}$ in the best fits 
of a simple SO(10) model, 
with the Wilson coefficient $C_7^{MSSM}$ of ({\em a}) the
same ({\em b}) the opposite sign as compared to $C_7^{SM}$. 
For better reference, the $\chi^2$ contour lines of figures 5a and 5b
are shown in the background of (a) and (b), respectively, as dotted lines.
}
\label{famu_F4c}
\end{figure}

When the future BNL experiment reduces the uncertainty on $a_\mu$ down to 
$\pm12\times10^{-10}$ at 90\%C.L., eq.(\ref{da_99}), the muon anomalous magnetic moment will 
undoubtfully turn into a powerful constraint on the MSSM analysis. It is already clear from figures 
\ref{famu_tb2_SUSY}, \ref{famu_tb20_SUSY}, and \ref{famu_F4c} that
it will be a major constraint for large and medium tan$\beta$ in the region $m_0 < 400-500$GeV.
Of course, the BNL result may drastically affect the MSSM analysis with any value of tan$\beta$ 
if the observed central value turns out to be below (or well above) the SM prediction. For such an outcome,
the muon anomalous magnetic moment will actually become a dominant constraint for the MSSM analysis
with the universal SUGRA-mediated SUSY breaking terms.

In the meantime, $a_\mu$ does not pose any new constraints on the MSSM analysis under the terms
considered in this work.

\vskip 0.3in
\noindent
 {\large {\bf Acknowledgments}}\\

I would like to thank Greg Anderson, Stuart Raby and Carlos Wagner 
for helpful discussions, various suggestions and support.

\newpage

%

\end{document}